# Smartphone Experiments in Physics Undergraduate Research

**ABSTRACT**
Smartphones and tablets are an integral part of our daily lives, and their capabilities extend well beyond communication and entertainment. With a broad choice of built-in sensors, using these mobile devices as experimental tools (MDETs) allows for a many different measurements, covering several fields of physics (mechanics, acoustics and waves; magnetism, optics, etc.).
Building on this development, the present contribution is about exploring the potential of MDETs in physics undergraduate research. Two examples related to acoustics (bottle Helmholtz resonator, singing glasses) will be discussed in detail, and four further possibilities are referred to as perspectives. Results of these student research projects are encouraging throughout (accuracy, agreement with theory), in many cases providing a basis for further improvements and insight. Additionally, it is argued that these examples provide a "proof of concept" that the use of smartphones for experimental projects can mobilize and stimulate among students the educational potential of "higher order thinking skills" (HOTs) widely discussed in the literature on undergraduate research, such as autonomy, curiosity, creativity, and others.

## 1 Introduction

Smartphones and tablets together with their built-in sensors provide compact and cost-efficient measurements for several physical quantities. Using these mobile devices as experimental tools (MDETs) allows for a broad range of measurements with a reasonable effort and in the same time satisfactory precision, and real-time data interpretation in both lab and everyday life situations. Recently, very complete reviews (Monteiro and Martí, 2022) and collections (Kuhn and Vogt, 2022) for MDETs in physics education have appeared, covering many fields of physics (mechanics, acoustics and waves; temperature and heat, electricity and magnetism, optics, modern physics, and astronomy).
Building on this development, the present work is about a "proof of concept" that the use of smartphones for experimental projects can mobilize and stimulate among students the educational potential of "higher order thinking skills" (HOTs) widely discussed in the literature on undergraduate research: autonomy, curiosity, creativity, problem solving, critical thinking (Zohar and Barzilai, 2015; Walsh et al., 2019; Murtonen and Balloo, 2019; Mieg et al., 2022). The case for such student research projects has been made in general e.g. by (Russel et al, 2007, National Academies of Sciences, Engineering, and Medicine, 2017; Monteiro and Martí, 2022; Abraham et al, 2022, Ahmad and Al-Thani, 2022), and the present contribution is about the role MDETS can play in that context.
Specifically, we report about the implementation of MDETs in the in the course "Physique du Quotidien" ("Physics of Everyday Life", very much in the sense of the "Flying circus of Physics" by J. Walker (2007, 2008)) held at the University of Geneva since several years. This implementation took place as follows: In an early lesson in the course, the potential of mobile devices as experimental tools was presented by co-authors LD and AM, providing some technological background, a series of motivating examples (from the PhD research project of Darmendrail, 2023, and other sources), and resources (literature, specific apps for different sensors, etc.). Moreover, literature for several areas of interest for the course topic were provided on the course website or by direct contact with the students (e.g. on music and physics). If students were interested in the use of MDETs (they could choose also other topics), they then autonomously thought about experimental projects to be carried out in the framework of course, and received feedback (feasibility) and hints for their work. Finally, the experimental students projects were presented in the course, and graded.
In the following, we discuss several student projects, where MDETs were used to carry out experiments and measure different parameters regarding of daily physics phenomena. The discussion is organized as follows: In sections 2 and 3, two student projects are presented more in detail, with (i) a brief account of the physics background (in some cases additional context is given, e.g. to increase student interest for teaching purposes); (ii) presentation of illustrative results; (iii) discussion of the results, in particular additional physics considerations, leading to improvements and further insights. For the other students projects, an overview is given in sect. 4, and a general discussion of the potential of MDETs in undergraduate education, in particular regarding HOTSs is given in sect. 5.

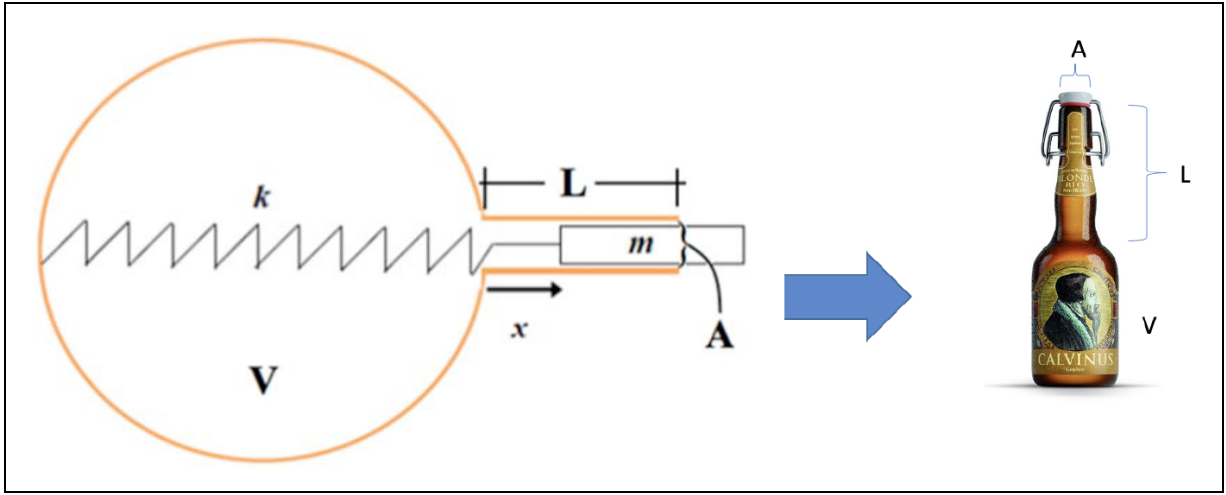

**Figure 1**: Helmholtz resonator model (left) and a real bottle with the respective dimensions. The air volume $V$ serves as "spring" (with spring constant $k$) which makes oscillate the "mass" ($m$, determined by $L$ and $A$).

## 2 Helmholtz resonators at a party[1]

In this student work the Helmholtz resonator model is applied to the well-known sound phenomenon produced by blowing over the mouth of a bottle (Swartz, 1994), and testing the theory for bottles of different size and shapes, and additionally for different filling volumes of the bottles.

### 2.1 Theory

Helmholtz resonances (HR) are at work with many acoustic phenomena in everyday life (musical instruments, such as fundamentals of guitar and violin (Fletcher & Rossing, 1998, 9.4, 10.5.2); the buffeting or booming sound in vehicle interiors (He & Yang, 2012); the modulation of the sizzling sound of a frying pan (Darmendrail & Müller, 2020), etc. The model of the Helmholtz resonator is shown in *Figure 1* (adapted from Matar & Welti, 2009): an air mass in the opening is coupled to the "spring" of compressible air within the cavity.

The Helmholtz resonance frequency is given by (Garrett, 2017; 8.5):

$$\nu_0 = \frac{c}{2\pi}\sqrt{\frac{A}{VL}} \qquad (1)$$

where $c$ is the sound velocity, $L$ the length of the neck, $A$ cross section of the neck, and $V$ the volume of the bottle.

### 2.2 Experiments

In total, 17 measurements with 5 different bottles and different filling volumes were carried out, considering different geometries and sizes.

**Figure 2** shows the sample of bottles used for the experiment. The procedure for the measurements is a follows:

- A hair dryer blows over the mouth of the bottle; this allows in general for a more stable and reproducible excitation of the Helmholtz resonance than blowing with the mouth.
- For each measurement the noise produced for the hair dryer is subtracted
- For a few cases (small volumes, high frequencies), the common method by blowing with the mouth is applied.
- For each bottle, additional measurements are taken by adding water, thus diminishing the volume and increasing the resonant frequency.
- Measurements taken with the application Spectrum Analyzer (Keuwlsoft, n.d.) on a Samsung Galaxy smartphone.

---

[1] This work was carried out by co-authors CB and LC.

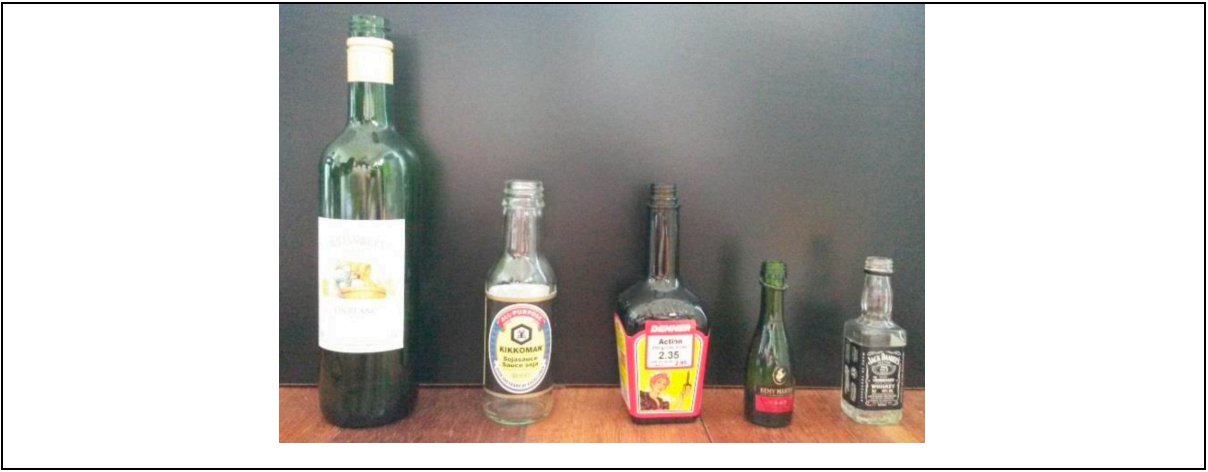

**Figure 2**: Bottles used in the experiment (from left to right: wine; soja; Maggi; Remy Martin; Jack Daniels).

Two examples of the measurements (wine and Jack Daniels bottle) are shown in **Figure 3**. As the wine bottle has a large volume, it was easy to carry out several measurements with different volumes (by filling in different amounts of water, see above). For the much smaller Jack Daniels bottle, only measurements for two volume values could be taken n a reliable way. In both cases, one sees the increase of frequency with the decrease of volume. We now turn to the quantitative results, and an appreciation of their quality.

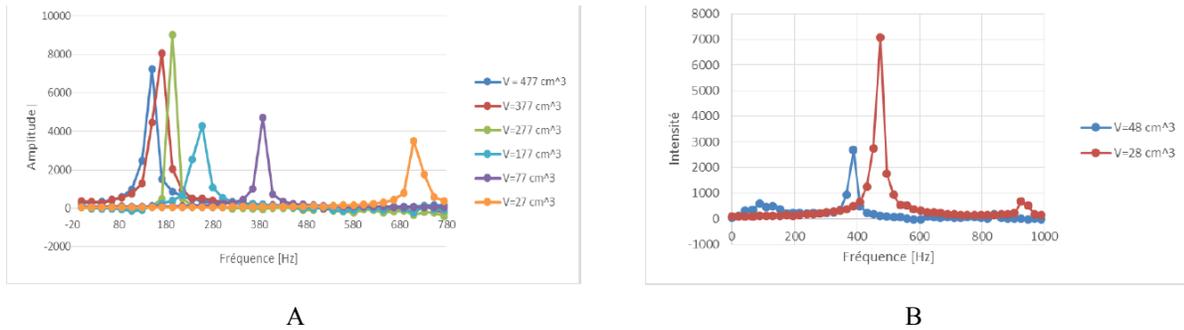

A                                                                                                  B

**Figure 3**: Acoustical frequency response for different volumes; y-axis: amplitude (arbitrary units); x-axis: frequency (Hz). A: wine bottle, B: Jack-Daniels bottle.

## 2.3 Results and comparison to theory

A display of the ratio of predicted values of $f_0$ (eq. 1) divided by experimental values allows for an overall comparison for the set of measurements taken (see Figure 4, left). Results show good overall general agreement for larger volumes, and acceptable agreement for smaller volumes (deviations still < 20%, except for one value). As the relative error goes like $|\Delta v_0 / v_0| = \tfrac{1}{2} \Delta V/V$ the larger deviations for smaller volumes are to be expected. Leaving out the one case with exceptionally large deviation, a plot of predicted values of $v_0$ (eq. 1) vs. experimental values (Figure 4, right) shows a very good agreement ($y \approx x$, as it should be, with a very high coefficient of determination $R^2 = 0.98$).

**Table 1**: Helmholtz frequencies for small Remy-Martin bottle, uncorrected and with shape correction (see text); $f_{0c}$ and $f_{0m}$ are the calculated and measured values, respectively, $\delta = (f_{0m}-f_{0c})/f_{0m}$ the relative error.

|             | $V\ [cm^3]$ | $v_{0c}\ [Hz]$ | $v_{0m}\ [Hz]$ | $v_{0c}/v_{0m}$ | $\delta$ |
|-------------|---:|---:|---:|---:|---:|
| uncorrected | 29 | 451 | 409 | 1,10 | −0,10 |
|             | 9  | 809 | 602 | 1,34 | −0,34 |
| corrected   |    | 402 | 409 | 0,98 | 0,02 |
|             |    | 722 | 602 | 1,20 | −0,20 |

The largest deviation occurs for the Remy Martin bottle, and inspection of its form (**Figure 2**) leads to a possible reason: there is a rather smooth transition between the neck and the main volume of bottle, so that the neck length cannot be defined unambiguously (this is also true for other bottle types, but to a lesser extent). For the values in

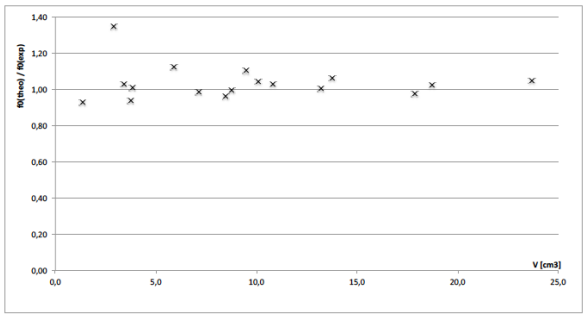
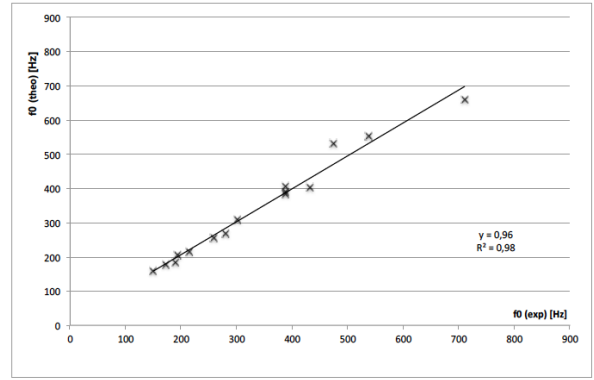

A                                          B

Figure 4, A: y-axis: ratio of predicted values (eg. (1)) and experimental values ($f_{0c}$ (calculated) / $f_{0m}$ (measured)); x-axis: volume $V$ (cm$^3$);
B): Helmholtz frequency, predicted values ($f_{0c}$; y-axis) vs. experimental values ($f_{0m}$; x-axis).

Figure 4, the neck was assumed to extend only to the point where the smooth widening begins. However, due to the smooth transition between neck and full width, the effective neck length is expected to be larger, the Helmholtz frequency thus to be lower (according to eq. (1)); this is a correction in the right direction, as the calculated frequencies lead to an overestimation. Taking the position in the middle of the transition region (i.e. 4.9 cm instead of 3.9 cm) leads indeed to improved values: The relative error goes down from 10% and 34% to 2% and 20%; (see **Table 1**). Moreover, the coefficient of determination analogous to Figure 4 is $R^2 = 0.96$, i.e. very high even including the case with the deviant shape now).

## 2.4 Discussion

This experiment is a good example for the use of MDETs for an experiment on a motivating, playful topic in the area of acoustics. Students showed creativity e.g. in establishing variation of test cases for a comparison with the Helmholtz resonator model, and also by replacing the usual excitation method by blowing with the mouth by the more practical and reliable method using a hair dryer. An analysis summarizing the measurements of the student project shows good to very good agreement with theory for all cases but one. Extending the student work further, a physical argument is presented to understand and correct the exceptional case, and indeed improved agreement is obtained. Other extensions of the HR model in a context of "everyday physics" are e.g. provided by Hirth et al. (2021). In sum, the example shows how MDETs can serve as basis for an undergraduate project with satisfactory results, and interesting possible extensions.

# 3  Singing glasses[2]

## 3.1 Theory

This topic is related to the well-known glass harp (of glass harmonica; also known as "verrillon" in French), where a glass is continuously exited by a (wetted) finger or another object like a violin bow. The glass vibratory behaviour can be described in terms of its geometry, density, and elasticity (or rigidity). French (1983) provides a derivation of a formula for the fundamental frequency accessible at undergraduate level.  For a cylindrical glass, one obtains

$$v_0 = \frac{1}{2\pi}\left(\frac{3Y}{5\rho_g}\right)^{1/2}\frac{a}{R^2}\left[1+\frac{4}{3}\left(\frac{R}{H}\right)^4\right]^{1/2}, \qquad (2)$$

(where $a$: glass thickness; $H$: glass height; $R$: glass radius, $Y$: Young's modulus, $\rho_g$: glass density)

## 3.2 Experiments

---

[2] This work was carried out by co-authors SB and DB.

### 3.2.1 *Fundamental frequencies for different glasses and comparison to theory*

In this student project, acoustical measurements were taken with a smartphone (app "spectroid" on Android devices) in order to find, among others, the fundamental frequency of several types of glasses, as well as for some further aspects (see 3.2.2 and 3.3).

For the fundamental frequencies measurements, ten different glasses were used (
Figure 5) in order to have a wide set of results to verify the applicability of eq. (2).

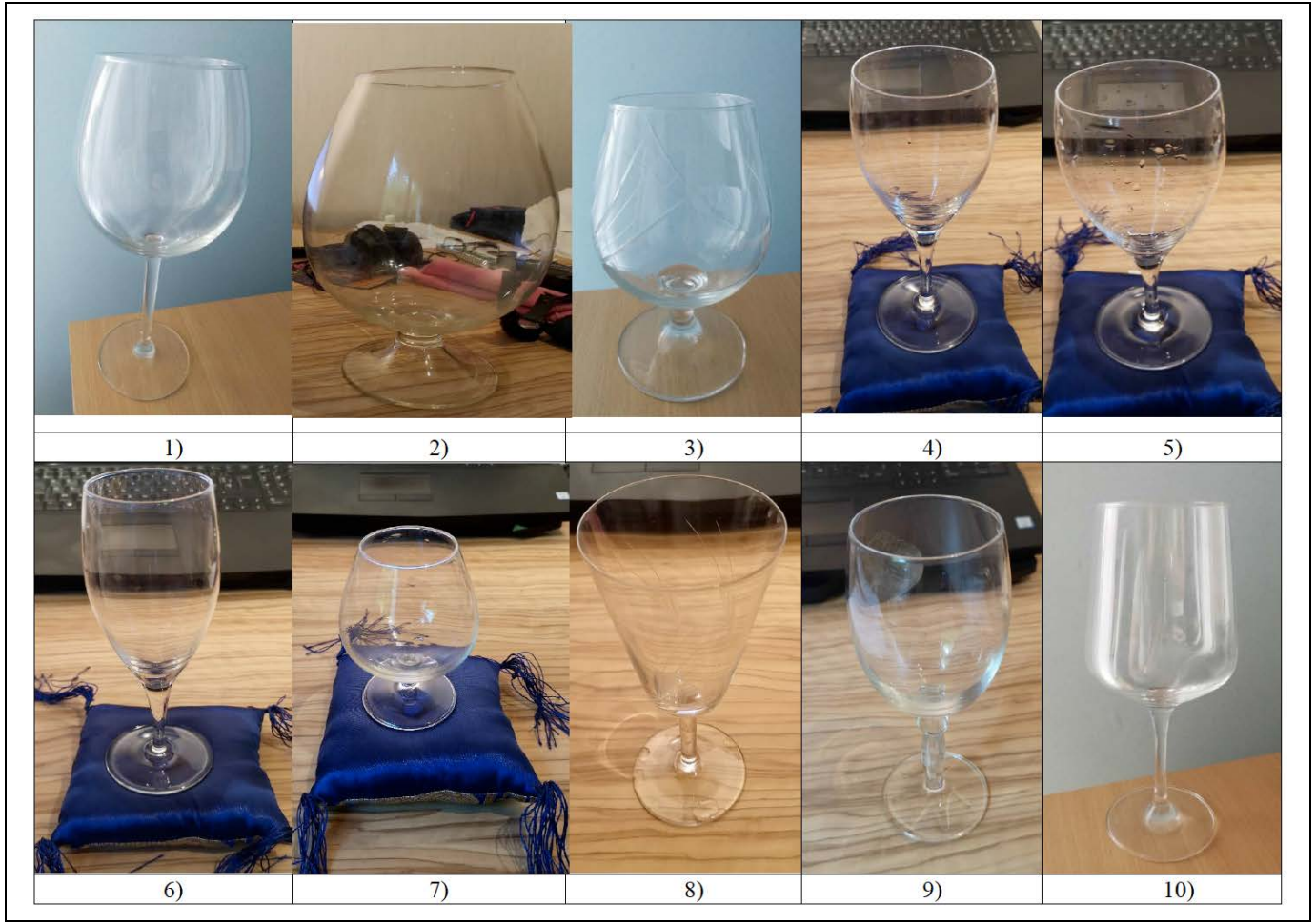

Figure 5: Glasses used in the experiment

Using typical values for Young's module and density of glass, **Table 2** shows the fundamental frequency results of ten different glasses considering their respective physical parameters and resulting relative error. The following statements can be made:

- for the cylindrical glass (no. 10) the agreement is very good ($\delta \approx 1\%$);
- the agreement is acceptable also for several glasses with other shapes (no. 1, 3, 4, 5; $\delta \approx 10\text{–}20\%$)[3];
- for other glasses larger deviations occur, in extreme cases up to $\delta \approx 50\%$ e.g. for a conical shape (no. 8).

---

[3] The agreement is surprisingly good even for glass no. 2, despite its strong deviation from a cylindrical shape ($\delta \approx 3\%$). It is possible that this decorative glass, intended for display rather than drinking, has different material properties that contribute to the observed agreement by chance. For this reason, we will not discuss this example further.

**Table 2**: Fundamental frequency results of ten different glasses, and relative error δ = ($v_{0m}$–$v_{0c}$)/$v_{0m}$): R is the value taken at the top of the glass; results are ordered by δ. A shape correction for non-cylindrical glasses allows to predict a corrected value for $v_0$ for bulbous shapes (see text).

| N° | a[cm] | H[cm] | R[cm] | shape cy(lindrical) bu(lbous) ta(pered) | $v_{0c}$ [Hz] (calculated) | $v_{0c}$ [Hz] (calculated, with shape correction) | $v_{0m}$ [Hz] (measured) | δ [%] | δ [%] (with shape correction) |
|---|---|---|---|---|---|---|---|---|---|
| 10 | 0.118 | 11.9 | 3.50 | cy | 534 |  | 538 | 1 |  |
| 2 | 0.392 | 20.0 | 6.25 | bu | 557 | –³ | 541 | -3 |  |
| 5 | 0.116 | 9.7 | 2.90 | ta | 764 |  | 699 | -9 |  |
| 3 | 0.190 | 10.0 | 3.35 | bu | 941 | 716 | 806 | -17 | 11 |
| 4 | 0.148 | 9.1 | 3.30 | ta | 758 |  | 938 | 19 |  |
| 1 | 0.180 | 11.0 | 3.65 | bu | 751 | 623 | 624 | -20 | 0,2 |
| 7 | 0.134 | 7.2 | 2.45 | bu | 1242 | 871 | 896 | -39 | 3 |
| 6 | 0.148 | 11.0 | 2.30 | ta | 1544 |  | 1099 | -40 |  |
| 9 | 0.136 | 8.2 | 3.45 | ta | 643 |  | 1090 | 41 |  |
| 8 | 0.110 | 8.5 | 4.00 | ta | 391 |  | 768 | 49 |  |

Going a step beyond the student analysis, one can distinguish bulbous shapes ("snifter" like, no. 1, 3, 7), and tapered shapes ("champagne flute" or white wine glass like, e.g. 4, 5, 6, 8, 9). For bulbous shapes, the glass has on average a larger radius than that at the top. It can easily be shown, that $v_0$ as a function of R according to eq. (2) is monotonously decreasing, thus the actual fundamental frequencies for these cases should be lower than predicted with R taken at the top, and this is indeed observed (see **Table 2**). Kasper and Vogt (2022) have shown that taking the arithmetic average of the radii at the top and at maximum diameter as a more accurate estimate of "effective" radius leads to satisfactory results. In our case this corrections brings the calculates values indeed closer to the measured ones (716, 623, 871 Hz for glass no. 3, 1, 7, respectively) and the relative error down to 10%, in the case of the strongest deviation from the cylindrical model for bulbous glasses (no. 7) from 39% to 2%. This can be seen as an exercise for the development of estimation and approximations skills, advocated by several authors as an important element in physics education under various terms like "back-of the-envelope calculations"; "guesstimation" (Swartz, 2003; Weinstein & Adam, 2008; Mahajan 2010), and related terms ("Order-of-Magnitude Reasoning", "Fermi questions"; Morrison, 1965; Weinstein, 2007). This way of reasoning is important as it provides estimates otherwise hard to obtain by precise calculations, e.g. to check the plausibility of a claim or a result. Morrison (1963) emphasizes its importance for "[t]he conception of experiments and the formation of theoretical hypotheses". This skill has been nicely put into the book title "Consider a Spherical Cow" (and its sister, the "Cylindrical cow" (Harte, 1988; Harte, 2022), see Figure 6; the "cylindrical cow" and the considerations related to it are in the present case the cylindrical shape of the glass, and expected corrections when departing from this simplifying assumption.

Let us now discuss the tapered shapes. Analogously to the case of bulbous shapes one could say that the average radius of the glass is smaller than that at the top, and then conclude that the actual fundamental frequencies for tapered shapes should be larger than those predicted with R taken at the top. This is true for several glasses, e.g. the cone shaped one (no. 8, also ok for no. 4 and 9), but not for all of them (no. 5, 6). A possible factor at work is that the precise form of the transition region between the wall and the stem of the glass will have an influence on the stiffness and hence on the frequency of the glass. No further investigation of the variations found for tapered glasses was carried out, and they have to be admitted as a limitation of the present approach.

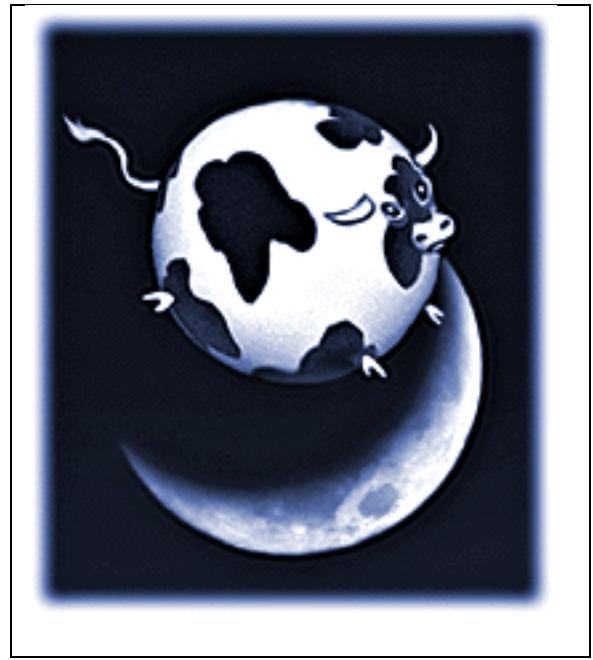

Figure 6: The "spherical cow", a metaphor for estimation in science (Harte, 1998)[4].

---
[4] Image source: Kallick, I. (1996), https://commons.wikimedia.org/wiki/File:Sphcow.jpg

### 3.2.2  *Additional experiments and observations*

As examples for the breadth of aspects which can be treated with MDETs in undergraduate teaching we briefly comment on further experiments carried out in that student project:

1. Higher modes: The student experiment successfully showed that a "singing glass" can be exited in a wide range of vibration modes (
Figure 7). The spectrogram taken with the app "spectroid" shows four distinctive peak frequencies with a clear fundamental at 639 Hz. With an elementary calculations it is readily found that the other three frequencies are not related as harmonics to this lowest frequency. As an example, the first harmonic of 639 Hz is 1278 Hz and the second harmonic is 1917 Hz, but the two next frequency peaks are 1324 Hz and 2531 Hz respectively and the differences are too high to be explained in terms measurement errors. Contrary to a misconception sometimes encountered, the fact that the higher modes of a complex vibrational system are not higher harmonics of the lowest frequency is well known in acoustics, see e.g. Chladni plates (Rossing, 1982) or the violin (Gough, 2007). For the wine glass, in particular, a detailed discussion in terms of a comparison to the flexural modes of a bell in given in (Rossing, 1994).

2. Excitation mechanism: Sounds from "singing glasses" can be excited by tapping, rubbing with a moist finger, leather mallet, or other objects, or by a bowing. The use of a bow allows to excite specific modes. By tilting the bow and applying different amounts of pressure, specific modes can be isolated as is shown in (
Figure 7, right). The results of the measurements with the violin bow shows that the frequencies of 1324 Hz and 2531 Hz correspond to higher modes of the wineglass.

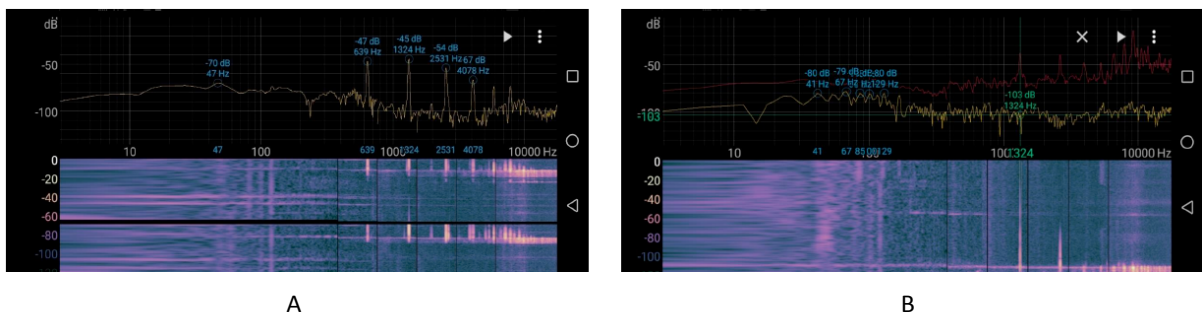

A                                                  B

Figure 7A: A wineglass exited in a wide range of vibration modes;
B: excitation with a violin bow allows excitation of specific modes.

## 3.3  Discussion

The example shows, that MDETs can be well used in an undergraduate project on the well-known phenomenon of "singing glasses". Students obtained good agreement between measurements and the calculated fundamental frequencies $f_0$ according to French (1983) for a cylindrical glass, and provided insight in some of the difficulties and limitations of the theoretical model theory and reality for other shapes. Extending the student work, satisfactory agreement can be obtained by applying a "shape correction" to bulbous glasses, based on a consideration of dependence of $f_0$ on the glass radius. For tapered shapes, the theoretical understanding of the phenomenon remains less satisfactory.

Additionally, it was shown that a range of further related interesting physical aspects can be investigated with MDETs in that student project, viz. higher modes and excitation mechanisms (3.2.2); furthermore partially filled glasses, important for the application in music, and well in accord with a model found in the literature (French, 1983); and musical instruments (glass harmonica, Rossing, 1994); Tibetan sound bowl, Terwagne and Bush, 2011).

Summarizing, this use case of MDETs also demonstrates (as the previous example) how they can be used to trigger experimental skills in an undergraduate project, and allow for interesting variations and further extensions.

## 4  Further examples

We briefly present, in form of an overview (**Table 3**), four other examples of the use of MDETs in student projects, three of them investigating forces in cars, airplanes, and a flight simulator, respectively, and a fourth the curvature and radius of the Earth as a foundational topic in the history of science:

- acceleration and friction coefficient of a car under different road conditions (Grand & Koziol, 2021);
- gravitational acceleration in a flight simulator (Mounzer & Sandoz, 2016);
- gravitational accelerations of parabolic ("zero g") flight in a small two-seat aircraft and in a model (home) experiment (Mileto & Martinez, 2019);
- a modern version of the determination of the Earth's radius according to Al-Biruni's method (Mopurgo & Zahler, 2019).

These examples show (see **Table 3** for an overview of the main features), how data and measurements illustrative and useful for physics learning can be obtained by MDETs, both qualitative (Mounzer & Sandoz, 2016; Mileto & Martinez, 2019) and quantitative (Grand & Koziol, 2021; Mopurgo & Zahler, 2019). Moreover, they show that the attentive, curious look of the physicist to observe and understand noteworthy phenomena of the world we live in is also well present in these student projects, as well as creativity and inventive character in the experimental approaches.

## 5  Conclusions: The potential of MDETs in undergraduate research projects

The main objective of this of this work was to explore the educational potential of MDETs in undergraduate research projects. Specifically, this took the form of experimental student projects in a course "Everyday Physics" (in the spirit of Walker 2007, 2008) at the University of Geneva over several years, based on an introduction by co-author LD. providing technological background, motivating examples, and resources (literature, apps), complemented by later exchanges (responding to questions, providing further literature, e.g. on music and physics, etc.).

Six use cases are discussed (see **Table 4** for a summary) here: two about acoustical phenomena or using acoustical methods (bottle Helmholtz resonator, singing glasses); see Dala Polla et al. (2023) for an additional example of this type; four additional ones in mechanics (car breaking distance, flight simulator, parabolic flight) and about a classical topic of astronomy (Earth radius). In all cases, elements of playful and stimulating approaches even for traditional topics are visible in these student approaches.

Moreover, extensions and further developments of the student work, leading to improvements and further insights have been presented (2 - 4). They show the heuristic, productive role of the student research, as it becomes a starting point for further investigations, a key element of the research process in general.

**Table 3:** Further examples of undergraduate projects using MDETs – overview for main features

| Sensor, app | Topic and exemplary results |
|---|---|
| | **Mechanics** |
| acceleration (e.g. by *SensorLog*) | 1) <u>Acceleration and friction coefficient of a car under different road conditions</u> (Grand & Koziol, 2021): dry (left), snow (right). Friction coefficients can be inferred from measured acceleration values by $\mu = a/g$, as treated in a standard mechanics course (see e.g. Halliday et al. 2020, sect. 5.8). In this way, average values of $\mu_{dry} = 0.89$ (0.02) and $\mu_{dry} = 0.25$ (0.03) were obtained in this student work, consistent with the ranges found in the literature (Logue, 1979; Ogawara, 2016). |
| | ![Two acceleration vs time plots showing car braking on dry road (left, $a_x = -0.87$, $\Delta t = 1.16$ [s]) and on snow (right, $a_x = -0.28$, $\Delta t = 2.40$ [s])] |
| acceleration (*Sparkvue*) | 2) <u>Gravitational acceleration in a flight simulator</u> (Lucerne Transport Museum; Mounzer & Sandoz, 2016) |
| | ![Plot of Acc Y and Acc XZ vs Temps [s] from 115 to 155 s, showing oscillating acceleration between roughly -12 and +12 m/s²] |

| | | |
|---|---|---|
| | 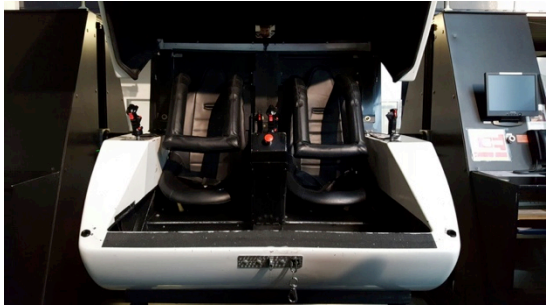 | Flight simulator (left), acceleration curves (right): During the turn (with the smartphone initially in horizontal position) one nicely sees the sign change of the z-acceleration (orange curve) as the device changes orientation with respect to the vertical, and the y- acceleration showing the same behaviour like the z acceleration with a phase delay of 90° |
| acceleration (e.g. by *SensorLog*) | 3) <u>Gravitational accelerations of parabolic ("zero g") flight</u> in a small two-seat aircraft and in a model (home) experiment (Mileto & Martinez, 2019). | |
| | 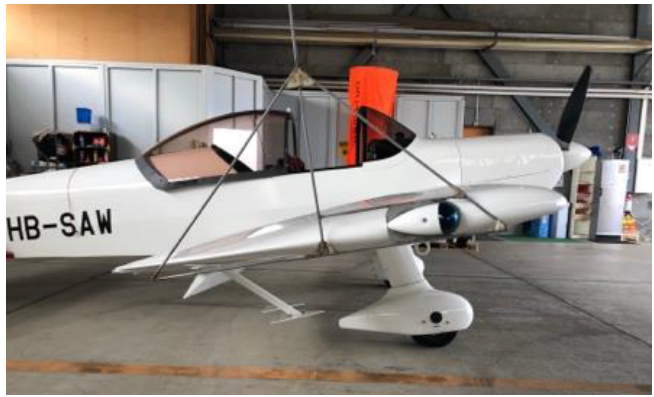 | 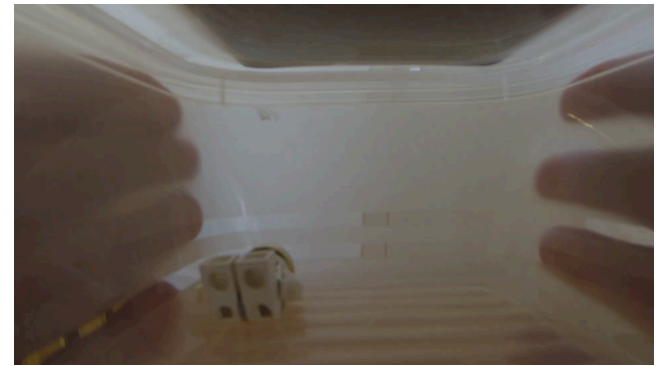 |
| | 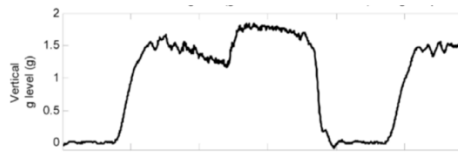 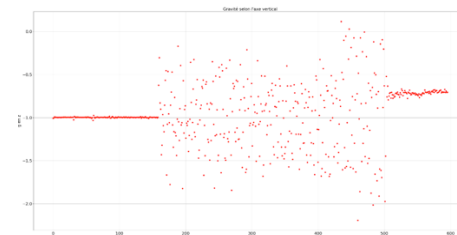 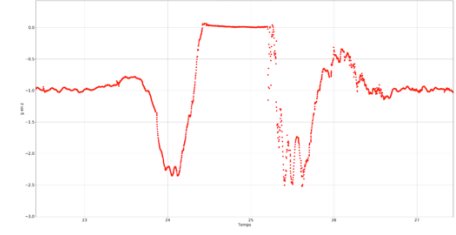 | |
| | Top left: Aircraft "Mudry cap 10" used for parabolic ("zero g") flight; top right: "flying box" used for zero g home experiment; Bottom left, top: zero g phases in a professional parabolic flight; middle: gravitational acceleration measures in "Mudry cap 10" aircraft; bottom right: gravitational acceleration measures in the home experiment. The initial idea was to demonstrate the zero g effect in a parabolic flight with a small aircraft. However, measurements where much two noisy (probably due to rough flight trajectory or/and vibrations in their aircraft). As an alternative, the zero g effect was measured in a home experiment with throwing a plastic box with a smartphone inside on a parabolic trajectory (inversion of z-axis due to upside-down attachment of the | |

| | | smartphone for practical reasons) |
|---|---|---|
| | | **Astronomy** |
| inclination (*Theodolite*) | 4) | A modern version of the determination of the Earth's radius according to Al-Biruni's method (Mopurgo, Zahler, 2019) Al-Biruni's method (Sparavigna, 2013) relates the angle $\alpha$ between the astronomical horizon and visual horizon at elevation $h$ to the radius of the Earth (see definition sketch below, right) $$R = \frac{h \cdot \cos(\alpha)}{1 - \cos(\alpha)}$$ In a modern version of this method, the observer at point A in the figure is in a plane. The angle $\alpha$ can be measured by the app "theodolite" (see below for an example for a measurement). The measurement yielded a value $\alpha = 2.69°$. Following the plane on the Flightradar24.com site the altitude was found as $h = 8330$ m. With these data, one obtains $R = 7600$ km, which is about 20% above the correct (averaged) value (6370 km). Taking into account atmospheric refraction, the value $R = 6300$ km is obtained, which is much more accurate (error $\approx$ 1%). |
| | 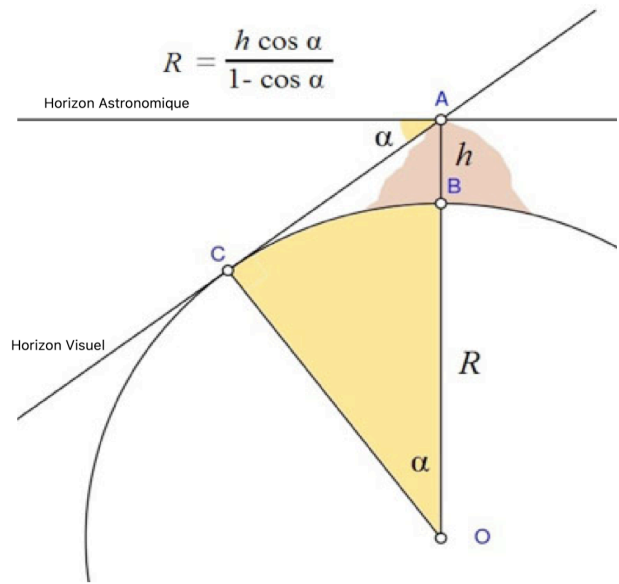 | 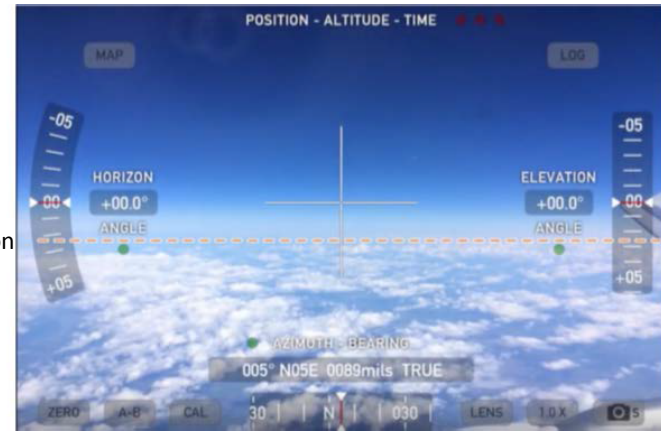 |

**Table 4**: Summary table: Undergraduate student experimental projects using MDETs (chronological order)

1) Acoustics: Helmholtz resonance "en soirée" (Béguin & Charosky, this contribution)
2) Acoustics: Singing glasses (Barro & Brouzet, this contribution)
3) Nonlinear Dynamics/ Mechanics: The Domino Effect (Dalla Pola et al., 2023)
4) Mechanics: Acceleration and friction coefficient of a car under different road conditions (Grand & Koziol, 2021)
5) Mechanics: gravitational accelerations of parabolic ("zero g") flight in a small two-seat aircraft and in a model (home) experiment (Mileto & Martinez, 2019)
6) Astronomy: A modern version of the determination of the Earth's radius according to Al-Biruni's method (Mopurgo & Zahler, 2019)
7) Mechanics: Gravitational acceleration in a flight simulator (Mounzer & Sandoz, 2016)

As one can see from these examples, a considerable breadth of topics can be treated with MDETs in undergraduate projects, from new demonstrations applications of basic physics concepts and effects to more advanced topics at the "higher end" of undergraduate physics (see Monteiro and Martí (2022) for a recent review and Kuhn and Vogt (2022) for a stimulating collection).

Beyond basic experimental competences (e.g. introducing variation, carrying out measurement series, considering errors, etc.) a specific interest of this work was on the potential of MDETs of mobilizing and strengthening higher order skills, and indeed one can see many instances of these in the student projects presented here: autonomy, in the choice of topics by the students, and in their independent work on the experiments with little guidance once a sufficient background had been provided (three projects were carried out during the Covid university closure beginning march 2020); curiosity, in "seeing" and investigating physics everywhere – e.g. cars, airplanes, a flight simulator in a museum (Swiss Museum of Transport, Lucerne), with bottles, glasses, and dominoes (Dalla Pola et al., 2023), etc; creativity, e.g. in the measurement series carried out, or the measurement methods based on the different smartphone sensors; problem solving, in overcoming experimental difficulties (e.g. using a hair dryer as constant source of excitation in the bottle Helmholtz resonator experiment); critical thinking, e.g. in considering limits of accuracy or of applicability of the theoretical models used (e.g. the correction for atmospheric refraction in the Earth radius measurement).[5]

Of course, there are limitations to these approaches: Regarding physics, it is clear that many MDET experiments have a limited accuracy and precision compared to standard equipment, and while their breadth of application in a "physics of everyday phenomena" setting is broad (see above), many measurements, in particular in a lab setting, cannot be carried out all. In fact, one has to take into account a trade-off of the availability and mobility of MDETs on the one hand, and more accurate, precise, and advanced lab experiments on the other hand. On the educational level, perhaps the aspect where students need most mentoring is the consideration of errors and uncertainties and critical assessment of results in an experiment (Hammack et al, 2017), a point to be considered when conceiving student research projects as part of the study program; note, however, that this is well-known as an important issue also from traditional labwork courses (Lippmann, 2003; Buffler et al, 2008; Priemer and Hellwig, 2018; Eshach and Kukliansky, 2018, Pols, 2021). Finally, regarding physics education research, an empirical study on the motivational and learning outcomes of the approach would be of interest, which was not in the scope of the present work.

By way of conclusion, and aware of the limitations, the experiences presented in this contribution provide a proof of concept for the use of mobile devices as experimental tools as a useful component in undergraduate physics education, specifically in the form of student research projects: by the breadth of the physics covered, the character of playfulness and stimulation, and competences to be learned, including higher-order thinking.

---

[5] Of course, these skills are strongly interrelated; e.g. one needs creativity for problem solving etc.

**Applications**

*Audio Recorder* (n.d.) https://f-droid.org/en/packages/com.github.axet.audiorecorder/

*Keuwlsoft Apps* (n.d.). https://www.keuwl.com/apps.html

*SensorLog* (n.d.). http://sensorlog.berndthomas.net/

*Sparkvue* (n.d.). PASCO Scientific. https://www.pasco.com/downloads/sparkvue

*Theodolite,* Augmented Reality Viewfinder for iPhone and iPad (n.d.). http://hunter.pairsite.com/theodolite/